\documentclass{PoS}
\usepackage{graphicx}
\usepackage{amssymb,amsmath,latexsym,enumerate}

\newcommand{\PDfrac}[2]{ \frac{\partial\,#1}{\partial\,#2} }

\title{On the causes and mechanisms of the\\
long-term variations in the GCR characteristics}

\ShortTitle{On the causes and mechanisms of the variations in the GCR characteristics}

\author{\speaker{M. Krainev}\\
Lebedev Physical Institute, Moscow, Russia\\
E-mail: \email{mkrainev46@mail.ru}}
\author{J. K\'ota\\
University of Arizona, Tucson, USA\\
E-mail: \email{kota@lpl.arizona.edu}}
\author{M.S. Potgieter\\
North-West University, Potchefstroom, South Africa\\
E-mail: \email{Marius.Potgieter@nwu.ac.za}}

\abstract{
We argue that the degree of understanding the causes and mechanisms of the long-term
variations (11-year and 22-year) in the galactic cosmic rays (GCR) characteristics
is still insufficient and to improve it we need new approaches and methods. For
the time being there is a long-lasting controversy on how these long-term variations, observed for more
than 50 years in the inner heliosphere, are formed. It is widely believed that the 11-year
variation is due entirely to the toroidal branch of solar activity (the area and
number of sunspots, the strength of the heliospheric magnetic field etc) because of
the diffusion, convection and adiabatic energy loss, while the much smaller 22-year
variation is caused by the particle drifts connected with the poloidal branch of
solar activity (the high-latitude solar magnetic fields). At the same time, both past
and more recent numerical simulations indicate that the contribution of particle drifts
could be significant for both 22- and 11-year variations in the GCR intensities.
However, even those who agree on the significant influence of drifts appear to have different perceptions
on the mechanisms of this influence.

In this paper, we present an analysis of the possible causes of the first point of
view (small role of drifts in the 11-year GCR variation) and the reasons why one can expect the significant contribution of the processes
connected with the poloidal branch of solar activity in both types of the long-term
variations of the GCR characteristics. Then we briefly discuss some numerical methods suggested in the past and recently
and the approaches and perspectives for the sought-for methods are considered.
}
\FullConference{The 34th International Cosmic Ray Conference,\\
                30 July- 6 August, 2015\\
                The Hague, The Netherlands}
\begin{document}

\section{Introduction}
\noindent The long-term (or more specifically the 11-year and 22-year) variations of the GCR
intensity and anisotropy, connected with the toroidal (or sunspot) and the poloidal
(high-latitude solar magnetic fields) branches of solar activity, have been studied
for more than fifty years. Many of the properties of GCRs are known from observations
mostly in the inner heliosphere. The transport partial differential equation (TPE)
serving as a theoretical basis for understanding and modeling these variations was formulated
about fifty years ago \cite{Parker_PhysRev_110_1445_1958,Krymsky_Geomagnetism_and_Aeronomy_4_977_1964,Parker_PSS_13_9_1965},
which was followed by extensive work on heliospheric models and
in-depth analysis of the TPE coefficients, describing the main processes
(diffusion, convection, adiabatic energy loss or gain, and particle drift) involved
in the modulation of GCR intensity. The systematic efforts of theoretical and numerical
modeling have resulted in a reasonable overall description of the observational
data \cite{Potgieter_LivReV_SolPhys_10_3_2013,Kota_SSR_176_391_2013}, although there are some questionable features in the simulations \cite{Krainev_etal_JPCS_409_012016_2013}.

However, the degree of understanding the causes and mechanisms of the long-term
variations in the GCR characteristics is still insufficient. As a cause of these variations
we mean the change of some heliospheric parameters (such as
solar wind velocity, strength of the magnetic field, its polarity etc.), connected with
one of two branches of solar activity or both of them, which gives rise to the GCR variation
under discussion. It should be stressed that to call the above change of the heliospheric
parameters the cause of the GCR variation it is not sufficient to demonstrate the regression
between them, but it is necessary to point out and understand how the mechanism of such influence
acts (see, e.g., \cite{Heise_Causal_Analysis_1975}). In studying the GCR variations theoretically,
we imply that their mechanisms are contained in the boundary-value problem for the GCR intensity
or, probably more practical, in the stochastic differential equations (SDEs) \cite{Zhang_ApJ_513_409_1999}
and in the coefficients of these equations (the diffusion tensor,
the solar wind and drift velocities and so on), used to model the intensity .
However, when one solves this theoretical basic problems - the boundary-value problem or SDEs -
to describe the observed
long-term variations of the GCR intensity, one does not yet obtain a full answer on the relative
importance of the different mechanisms and how it changes in different times, and in different regions
of energy and space. To obtain these answers one needs some additional means and approaches for
treating the theoretical basis for the GCR modulation.

For the time being there is a long-lasting controversy on the main causes and mechanisms of the long-term
variations of the GCR characteristics. There is a widespread view (especially among the observers),
that the 11-year variation, $J_{11}(t)$, is almost entirely due to the sunspot cycle (it is often even called
the sunspot cycle, $J_{ss}(t)$, in GCR intensity), while the much smaller 22-year
variation, $J_{22}(t)$, is connected with particle drifts. As to the GCR anisotropy, its 22-year varying component, W-wave,
was the first observational fact
that the GCRs are sensitive to the polarity of the high latitude solar magnetic fields \cite{Forbush_JGR_72_4937_1967,Forbush_JGR_74_3451_1969}. And  the first drift model \cite{Levy_14ICRC_4_1215_1975} was devoted to description of this W-wave.
However, the second component of the anisotropy, 11-year varying V-wave, was thought to be due entirely to the sunspot cycle.

On the other hand, the
active role of drifts in forming the 11-year variation of the GCR intensity followed
from several model calculations (mostly using first generation drift models)
\cite{Jokipii_Kopriva_ApJ_234_384_1979,Jokipii_Thomas_ApJ_243_1115_1981,Kota_Jokipii_ApJ_265_573_1983, Potgieter_Moraal_ApJ_294_425_1985,
LeRoux_Potgieter_ApJ_442_847_1995}.
The contribution of different mechanisms to the solar modulation of the GCR
intensity was also scrutinized in several recent papers (see
\cite{Bobik_etal_ApJ_745_132_2012,Kalinin_etal_33ICRC_297_2013,Krainev_etal_33ICRC_305_2013,Zhao_etal_JGR_119_1493_2014, Potgieter_etal_SolPhys_289_391_2014}).
Besides, recently a new numerical tool was suggested to disentangle the effects of all the
individual mechanisms from the full calculated intensity
\cite{Krainev_Kalinin_JoPCS_409_012155_2013,Krainev_JoPCS_inpress_2015}.

In this paper we first consider the reasons for the different points of view on the
causes and mechanisms of the long-term variations of the GCR intensity and
anisotropy, then discuss some suggestions of the approaches and methods for
improving the quantitative understanding of these variations. The possible perspectives
for the sought-for methods are also considered.

\section{On different points of view}
\noindent There are several causes of the widespread belief that the 11-year variations in the GCR
intensity and anisotropy are almost entirely due to the sunspot cycle The main cause is that (1) there is an overall
anticorrelation between the GCR characteristics and different solar and heliospheric
indices changing in phase with the solar activity cycle (the sunspot area, the
strength of the regular heliospheric magnetic field (HMF) and its inhomogeneities,
etc.). Besides it is important that (2) in the first 20 years (1955-1975) it was
believed that there was only the 11-year sunspot cycle in the heliospheric parameters and there
was not any dominating HMF polarity $A$ (the sign the regular HMF radial component $B_r^{hmf}$
in the N-hemisphere). After that $J_{22}(t)$ is considered as small
variation with respect to $J_{11}(t)$ as (3) the Earth is not in a favorable position
to observe the drift effects \cite{Potgieter_Moraal_ApJ_294_425_1985}.
Furthermore, it is often presumed (wrongly) that (4) the deciding
cause of a powerful 11-year cycle of GCR flux must be a powerful sunspot
cycle alone while the change of the polarity of the HMF, together with
the accompanying change of the heliospheric current sheet (HCS) between low and high tilt can
only results in small changes in the intensity of GCRs. Finally, it is also often presumed that (5)
the 11- and 22-year cycles in the GCR intensity
should be caused by the causes of the same periods. We should admit that the significant role of drift
is still questioned from observations even during the minima of solar cycles
\cite{Cliver_etal_SSR_176_3-19_2013} (see the discussion in \cite{Potgieter_etal_SolPhys_289_391_2014}).

The alternative view, that both 11-year- and 22-year variations in the GCR characteristics
are caused by the combined
actions of the sunspot cycle in the heliospheric characteristics and change of the
HMF polarity together with the HCS tilt, the latter being important for both GCR variations, is based
on the assurance that (1) the change of the dominating HMF polarity $A$ with 22-year cycle is a global process in the whole heliosphere and (2) there are mechanisms of the 11-year variation in the GCR intensity
forming as the 2-nd harmonic connected with the cause, changing with the 22-year period.
Among these mechanisms we can mention:
i) modulation of the 22-year mechanism with an 11-year period. For example, the drift velocity
depends on the HMF polarity (changing with 22-year period), and on the HMF strength and the
form of the HCS (both changing with 11-year period);
ii) simultaneous change of two factors in the 22-year mechanism. For example, the direction of
the drift velocity and the main channel of particle`s arrival are changing synchronously;
iii) two 22-year mechanisms acting in tandem. For example, if the GCR intensity is first
modulated by some hypothetical mechanism external to the part of heliosphere inside the termination shock
and then by drifts in this part, acting in phase with each other
(both as $\sin(\omega_{22}\cdot t)$), then some part of the modulated intensity changes as
$\sin^2(\omega_{22}\cdot t)=(1-\cos(\omega_{11}\cdot t))/2$, where $\omega_{22}=2\pi/22y$, $\omega_{11}=2\pi/11y$.

\section{Some numerical tools and approaches}
\noindent To discuss the numerical methods for studying the role of different mechanisms in the long--term variations of the GCR characteristics
we should first formulate the
boundary-value problem they deal with, first introduced in \cite{Parker_PhysRev_110_1445_1958,Krymsky_Geomagnetism_and_Aeronomy_4_977_1964,Parker_PSS_13_9_1965}.
Usually instead of GCR intensity
$J(\vec{r},T,t)$ the boundary--value problem and TPE are formulated for the phase-space
distribution function $f(\vec{r},p,t)=J(\vec{r},T,t)/p^2$, where $p$ is the momentum
of particles. For the stationary case the TPE balances the divergences of
the diffusion, drift and convection fluxes in space and that due to adiabatic cooling in
the momentum space. For the case of axial symmetry in a spherical system of
coordinates the boundary--value problem, i.e., the transport equation with boundary
and "initial" conditions (which may also be written somewhat differently), appears as follows:

\begin{align}
-\PDfrac{f}{t}=\underbrace{- \nabla\cdot(K\cdot\nabla f)}_{\mbox{diffusion}} +
\underbrace{{\vec{V}}^{sw}\cdot\nabla f - \frac{\nabla\cdot{\vec{V}}^{sw}}3p\frac{\partial
f}{\partial p}}_{\mbox{convection+adiabatic loss}} +
\underbrace{{\vec{V}}^{dr}\cdot\nabla f}_{\mbox{drift}}&=0,\label{TPE_f_General}
\end{align}
\begin{alignat}{3}
\left.\PDfrac{f}{r}\right|_{r=r_{min}}&=0, \qquad \left. f\right|_{r=r_{max}}&=&f_{nm}(p),\qquad \left.\PDfrac{f}{\vartheta}\right|_{\vartheta=0,\pi}=0\label{Boundary_conditions}\\
\left.f\right|_{p=p_{max}}&=f_{nm}(p_{max}),& \label{Initial_condition}
\end{alignat}

\noindent where $\vec{V}^{sw}$, $\vec{V}^{dr}$ and $K^s$ are the solar wind and
magnetic drift velocities, and symmetric part of the diffusion tensor, respectively. Besides, $f_{nm}(p)$
is the distribution function corresponding to nonmodulated GCR intensity and
$r_{min}$, $r_{max}$, $p_{max}$ are the radii of the inner and outer boundaries of
the modulation region and the momentum above which there is no modulation.

Besides the GCR intensity we are also interested in the long-term variations in the GCR anisotropy $\vec\xi=3\vec S/(4\pi p^2vf)$,
connected with the differential streaming $\vec S$. The latter can be easily calculated after solving the boundary-value problem
Eqs. (\ref{TPE_f_General}-\ref{Initial_condition}),
$\vec{S}=4\pi p^2\left(C_{CG}f{\vec{V}}^{sw}-K\cdot \nabla f\right)$,
where $C_{CG}=-1/3\partial(\ln f)/\partial(\ln p)$, $K$ and $v$ are the Compton-Getting factor, the full diffusion tensor,
and the particle's velocity, respectively. In this paper we shall concentrate on the influence of the particle drift on the GCR intensity, as the role of drifts in forming the 11-year V-component in the anisotropy depends on their role in the 11-year variation of the GCR intensity.

The importance of the particle drifts for
the GCR intensity was proposed in \cite{Jokipii_Levy_Hubbard_ApJ_213_861_1977} soon after the HMF model appeared of two unipolar
magnetic ``hemispheres'' divided by the thin global HCS \cite{Schulz_ASS_24_371_1973}. The
models \cite{Jokipii_Kopriva_ApJ_234_384_1979,Jokipii_Thomas_ApJ_243_1115_1981,Kota_Jokipii_ApJ_265_573_1983,Potgieter_Moraal_ApJ_294_425_1985}
were the first ones getting both the 11-year- and 22-year long-term GCR variations from changing only the HCS tilt.
These models had very specific predictions, for instance,
marked differences in the GCR intensity time-profiles in consecutive solar cycles (plateau
vs peaked shape in $A>0$ and $A<0$ cycles for positively charged particles, respectively), as well as differences
in the radial and latitudinal gradients of GCRs.
These features were confirmed by observations \cite{Ahluwalia_16ICRC_12_182_1979,Cummings_Stone_Webber_GRL_14_174_1987}.
Their magnitude, however, tend to be smaller than prediction obtained by purely drift-dominated models
\cite{Potgieter_leRoux_Burger_JGR_94_2323_1989}. Some suppression of the drifts and modification of the TPE coefficients
(the compound approach, \cite{Ferreira_Potgiter_ApJ_603_744_2004}) helps in describing many of the observed features in the GCR behavior
quantitatively.

Already the first full drift calculations \cite{Jokipii_Kopriva_ApJ_234_384_1979,Potgieter_Moraal_ApJ_294_425_1985} showed that drift tended to reduce modulation by significant factor for both polarity cases and now it is a common knowledge for the modeling community (see \cite{Kota_SSR_176_391_2013,Manuel_etal_SP_289_2207_2014}). Recently in \cite{Kalinin_Krainev_JoPCS_409_012156_2013}, first, the parameters of the models were
chosen in such a way that the calculations basically described the latitude, radial
and energy dependencies of the observations for the HMF polarities,
respectively, in 1987 and 1997 solar minima. Then
the calculations were repeated with the same parameters but without the drift, that
is omitting the drift term in Eq. (\ref{TPE_f_General}). The calculated intensity almost
everywhere decreased by a factor of 3-5 when compared with the intensities for both
$A>0$ and $A<0$ periods. In \cite{Krainev_etal_33ICRC_305_2013} it was shown
that for the GCR intensity near the Earth the same was true for all other phases of solar cycle except the years near
solar maxima. So the conclusion was made that in the models used, similar to  \cite{Jokipii_Kopriva_ApJ_234_384_1979,Potgieter_Moraal_ApJ_294_425_1985,Manuel_etal_SP_289_2207_2014}, the drift
contribution was large for both polarities and it was
emphasizes that particle drifts result in the significant part of the 11-year variation of the GCR intensity in the whole heliosphere.

%
In \cite{Potgieter_etal_SolPhys_289_391_2014} devoted to modeling the unusual 23/24 minimum between solar cycles 23 and 24 (2009)
the extent to which diffusion and particle drifts contributed to the total observed modulation
from 2006 to 2009 was also investigated. For this purpose the 3D stationary TPE (\ref{TPE_f_General}) was solved by the finite-difference method. Besides the run with changes of all modulating parameters, the authors made the multiple runs each time changing only one of the factors,
(e.g., the HCS tilt), while every other modulation parameter was kept unchanged.
We shall not discuss the conclusions made by the authors, as here we are interested only in their approach to find the contribution of individual processes. In \cite{Zhao_etal_JGR_119_1493_2014} also dealing with the unusual 23/24 solar minimum approximately the same method (the multiple runs each time changing only one of the factors) was used also to find the relative importance of different modulating factors.
The main difference from the method used in \cite{Potgieter_etal_SolPhys_289_391_2014} is that in \cite{Zhao_etal_JGR_119_1493_2014} the time-backward Markov stochastic process method, proposed in \cite{Zhang_ApJ_510_715_1999,Zhang_ApJ_513_409_1999}, was used to solve the 3D SDEs.
The questions of how the drift and other processes influence the propagation times and energy losses in the heliosphere were addressed in \cite{Strauss_etal_JGR_116_A12105_2011} also using SDE approach. The same method was used in \cite{Bobik_etal_ApJ_745_132_2012} also changing one factor at a time (switching on and off the drift term and modifying the diffusion tensor) in multiple runs.

In \cite{Krainev_Kalinin_JoPCS_409_012155_2013} the
method was suggested to decompose the calculated GCR intensity into the partial
intensities connected with the main physical processes (the diffusion, convection,
adiabatic cooling and magnetic drift) of the solar modulation, $J=J_p^{diff}+J_p^{conv}+J_p^{adiab}+J_p^{drift}$.
In more details the method is described in \cite{Krainev_JoPCS_inpress_2015}, where
it is also demonstrated how it can be used for studying the mechanisms of the GCR
intensity modulation and the heliospheric structure during the solar minima with
opposite HMF polarities ($A>0$, $A<0$) and without drift effects ($A=0$).

%
\section{Discussion and conclusions}
\noindent It looks like there is a hint of consensus between the modelers on the significance of particle drifts for both 11-year and 22-year variations in the GCR intensity. So when one discusses the observed drift effects, not only the difference between the GCR characteristics for $A>0$ and $A<0$ periods should be taken into account, but also the probable significant contribution of drifts in these characteristics for both polarities. As it is impossible to switch off particle drift in nature, estimating this contribution from the observations is a difficult task. Probably, drifts should also strongly influence the 11-year V-wave in the GCR anisotropy.

However, what we need are the numerical means for the quantitative study how the different modulation mechanisms form the long-term variations in the GCR intensity.
Note that in all models \cite{Potgieter_etal_SolPhys_289_391_2014,Zhao_etal_JGR_119_1493_2014,Bobik_etal_ApJ_745_132_2012}, studying the contribution of different mechanisms by the method of multiple runs changing only one of the factors at a time, the overall distribution of the intensity gradients  for each such run is not the same as it is in the run with all the modulating factors changing in time simultaneously.
This difference in the intensity gradients can be very significant and so the diffusion, drift and convection of the particles, that is all the processes depending on the gradients, carry out differently in each of the runs. On the other hand, the decomposition of the calculated GCR intensity into the partial
intensities connected with the main physical processes in \cite{Krainev_Kalinin_JoPCS_409_012155_2013,Krainev_etal_33ICRC_305_2013,Krainev_JoPCS_inpress_2015} is carried out in the run with all the modulating factors changing in time simultaneously, not disturbing the overall distribution of the gradients and hence the modulating processes.
Probably, it is one of the causes of different results on the relative contribution of the drift and diffusion to the GCR intensity near the Earth during the unusual solar minimum 23/24 which were obtained in \cite{Potgieter_etal_SolPhys_289_391_2014} and \cite{Krainev_etal_33ICRC_305_2013}. According to \cite{Potgieter_etal_SolPhys_289_391_2014} the relative contribution of the drift and diffusion is approximately the same, while in \cite{Krainev_etal_33ICRC_305_2013} the drift partial intensity is always much smaller than the diffusion partial intensity the latter being almost compensated by the large negative partial intensity connected with the adiabatic energy losses. It should be said that the results of \cite{Krainev_etal_33ICRC_305_2013} strongly depend on the models used (too simple and outdated in \cite{Kalinin_Krainev_JoPCS_409_012156_2013,Krainev_etal_33ICRC_305_2013,Krainev_Kalinin_JoPCS_409_012155_2013,Krainev_JoPCS_inpress_2015}) and there are unresolved questions in the method of partial intensities  \cite{Krainev_Kalinin_JoPCS_409_012155_2013,Krainev_JoPCS_inpress_2015}. However, what we are looking for is the numerical instrument, which can be applied to any model.

As up to now there is no accord on the ways how the long-term variations in the GCR characteristics are formed, even those who agree on the significant influence of drifts have different perceptions on the mechanisms of this influence.
{\underline{The first picture}} is that the GCR modulation is directly associated with the energy loss,
and the latter depends on the path and dwelling time of GCRs in the heliosphere, these quantities being determined by the interplay of the
processes involved. The drifts reduce modulation for both $A>0$ and $A<0$ periods, since drifts assist GCR particles to penetrate
into the inner heliosphere faster in both polarity cases (either along the pole or at the HCS) \cite{Kota_16ICRC_3_13_1979,Tverskoy_ASR_1_5_1981,Strauss_etal_JGR_116_A12105_2011}.
{\underline{Another picture}} arising from the consideration of the energy dependence of the gradients and the partial GCR intensities in the whole heliosphere \cite{Krainev_JoPCS_inpress_2015},
is that the changes in the intensity distribution start at the highest energy in the outer heliosphere and accumulate with decreasing energy and radial distance. The main result is the change of the intensity gradients which influences the diffusion and other fluxes in such a way that the GCR intensity increases for both HMF polarities.

The consensus in both pictures is that drift becomes more important and more apparent at larger helioradii and higher energies, and the global modulation is largely decided in the remote heliosphere (beyond tens of AU). Besides the main role of the energy loss in the first picture agrees with the conclusion in \cite{Krainev_etal_33ICRC_305_2013} that the largest negative partial intensity is $J_p^{adiab}$ (at least for cases considered there). It is also probable that the main difference in two above pictures is that in the first case we think of individual particles, while in the second picture we discuss the formation of the particle distribution: the two differs, but at the end could come with the same result.

We also hope that both the finite-difference and stochastic methods can help.
Moreover, probably the method to improve the quantitative understanding of the long-term variations in the GCR characteristics consists in the synthesis of the two approaches.

\section{Acknowledgments}
\noindent MK thanks the Russian Foundation for Basic Research (grant 13-02-00585) for the financial support.
MSP expresses his gratitude for the partial funding granted by the South African National Research Foundation (NRF) under the Incentive and Competitive Grants for Rated Researchers. Many features of this paper were formed during the preparation of the proposal to some grant.
The authors thank all the members of the team for their cooperation if this effort.

\end{document}